\documentclass[final,1p,times]{elsarticle} 
\usepackage{graphicx} 
\usepackage{amssymb} 
\usepackage{amsthm} 
\usepackage{lineno} 

\journal{Nuclear Physics A} 
\begin{document} 

\begin{frontmatter} 

\title{Formation Time of QGP from Thermal Photon Elliptic Flow}

\author{Rupa Chatterjee and Dinesh K. Srivastava}

\address{Variable Energy Cyclotron Centre, 1/AF Bidhan Nagar, 
Kolkata 700 064, India}

\begin{abstract} 
We show that the transverse momentum dependent elliptic 
flow $v_2(p_T)$ of thermal photons is quite sensitive to 
the initial formation time ($\tau_0$) of Quark Gluon 
Plasma (QGP) for semi-central collision of gold 
nuclei at RHIC~\cite{tau}. A smaller value of the formation 
time or a larger initial temperature leads to a significant 
increase in the thermal photon radiation from QGP phase, 
which has a smaller $v_2$. The elliptic flow of thermal 
photon  is dominated by the contribution from the quark 
matter at intermediate and high $p_T$ range and as a 
result sum $v_2$ decreases with smaller $\tau_0$ for $p_T 
\ge 1.5$ GeV. On the other hand we find that the elliptic 
flow parameter for hadrons depends only marginally on the 
value of $\tau_0$.
\end{abstract}

\end{frontmatter}

\section{Introduction}
Heavy ion collisions at relativistic energies lead to 
formation of Quark-Gluon Plasma, a deconfined 
novel state of quarks and gluons in local thermal equilibrium.
Interesting results of jet quenching due to parton energy 
loss in the medium~\cite{jetq} and elliptic flow of identified 
particles~\cite{v2} at the Relativistic Heavy Ion Collider (RHIC) 
at Brookhaven National Lab have provided significant 
evidence of the formation of QGP. However, one of the most important 
issues in this study, the initial formation time of the plasma, 
or the onset of collectivity and thermalization in the system, 
beyond which the powerful method of hydrodynamics can be used 
to describe its evolution is still not known precisely. 

In a very simple treatment it is assumed that, the partons 
produced in the collisions have an average energy $\langle E 
\rangle$, and thus their formation time $\tau_0 \sim 1/\langle 
E \rangle$. Looking at the various treatments and perhaps 
complementary models available in the literature, a direct 
measurement of the formation time would be very desirable. 
We show that the elliptic flow of thermal photons using ideal 
hydrodynamics can be very useful to estimate the value of 
$\tau_0$ accurately~\cite{tau}. 
\begin{figure}[t]
\begin{minipage}{15pc}
\includegraphics[width=13pc,angle=-90]{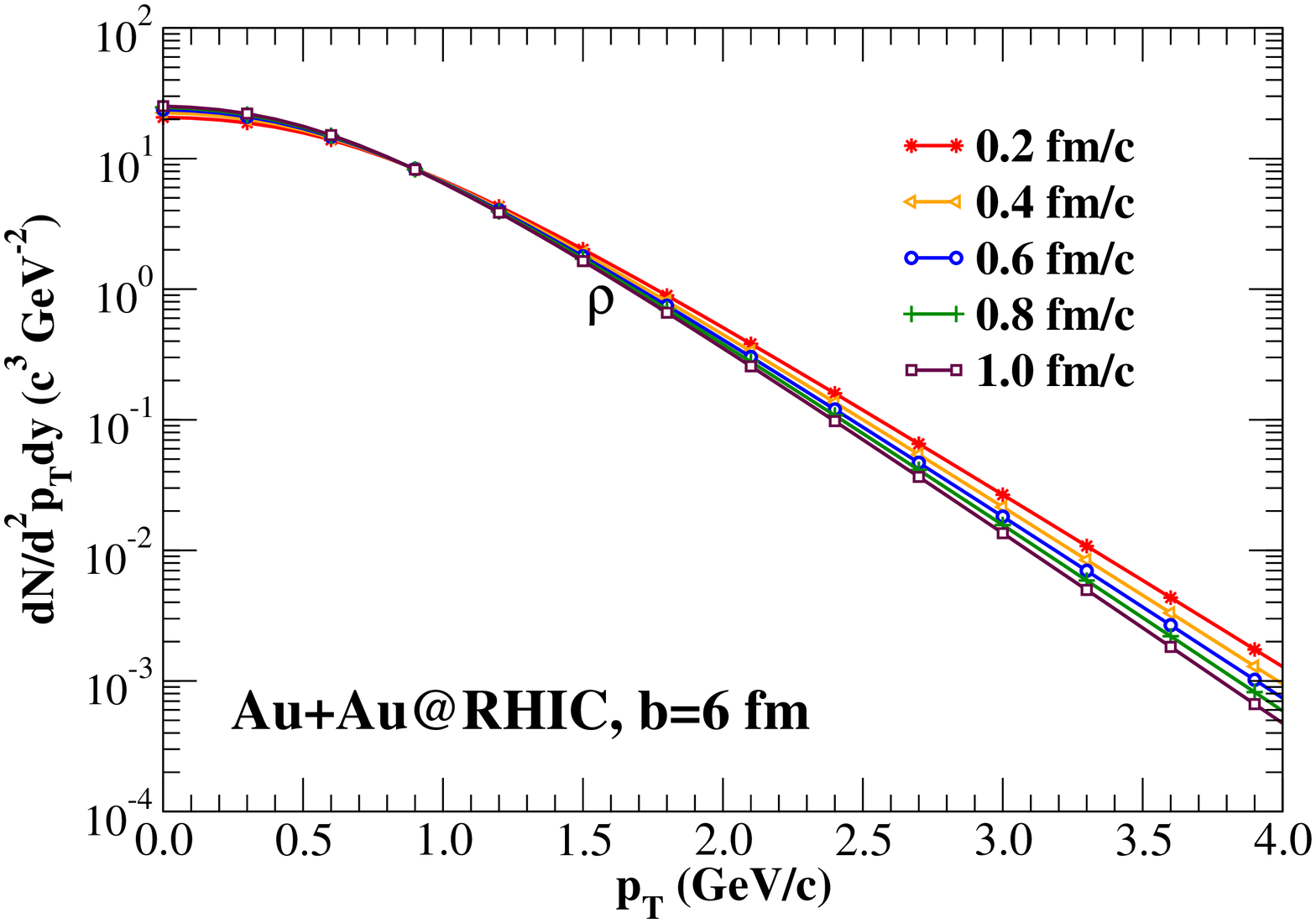}
\end{minipage}\hspace{1pc}%
\begin{minipage}{15pc}
\includegraphics[width=13pc,angle=-90]{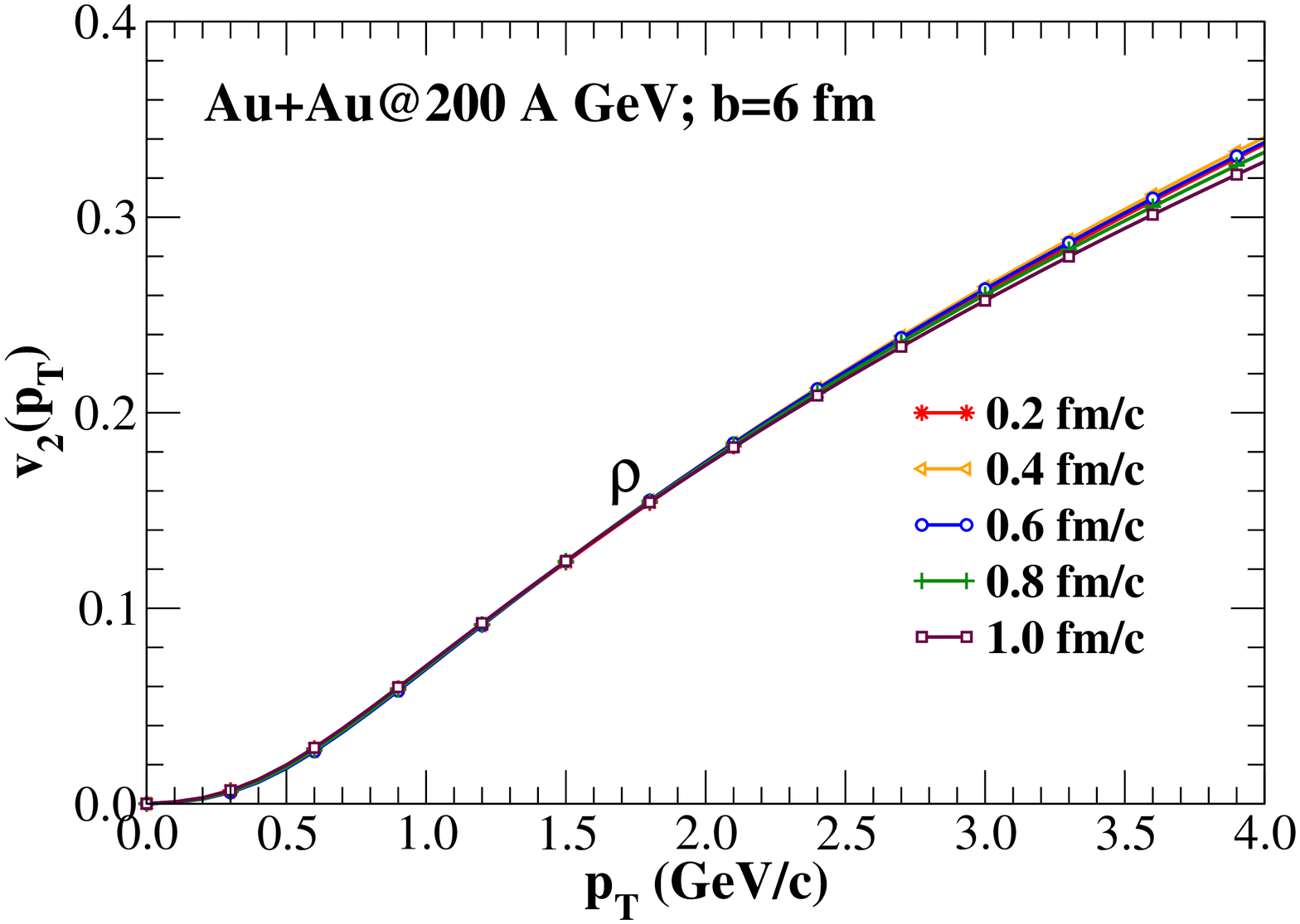}
\end{minipage} 
\label{rho}
\caption{$p_T$ spectra [left panel] and elliptic flow [right panel] 
of $\rho$ mesons considering  different initial formation time 
$\tau_0$ of the plasma for 200A GeV Au+Au collisions at RHIC and 
at b=6 fm.}
\end{figure}
\section{$\tau_0$ sensitivity of the elliptic flow parameter at RHIC}
It is quite well accepted that, photons are one of the most 
efficient probes to explore the properties of the hot and 
dense system produced in the collision of heavy nuclei at 
relativistic energies. Being electromagnetic in nature, 
they do not suffer any final state interactions and carry 
undistorted information about the circumstances of their 
production directly to the detector. Also, photons are 
emitted throughout the life time of the evolving system, 
whereas hadrons are emitted only from the surface of freeze-out.

In our earlier interesting work on elliptic flow of thermal photons 
at RHIC, we reported that the $v_2(p_T)$ of thermal photons using 
ideal hydrodynamics exhibits a completely different nature 
compared to the elliptic flow of hadrons at intermediate 
and large $p_T$~\cite{cfhs}. Thermal photon $v_2$ from 
quark matter is very small at large $p_T$ or early times, it 
gradually rises with smaller values of $p_T$ and then falls 
again as $p_T \rightarrow$ 0 [see Fig. 3 of Ref.~\cite{cfhs}]. 
The contribution to $v_2$ from hadronic matter photons rises 
monotonically with $p_T$, which is similar to the elliptic flow 
of hadrons predicted by hydrodynamics. 
The sum $v_2$ tracks the $v_2 (p_T)$ from quark matter at 
large $p_T$ (inspite of very large $v_2$ values from hadronic 
matter) as the quark matter radiation dominates the spectra 
beyond $p_T$ value of about 1 GeV/$c$.  Thus, the thermal 
photon $v_2$ at large $p_T$ reflects the momentum anisotropies 
of the partons produced at early times, soon after the collisions.

We follow the same treatment and initial conditions as used in 
Ref.~\cite{cfhs} to calculate the $p_T$ spectra and elliptic 
flow of thermal photons at different $\tau_0$ for semi-central 
collision of Au nuclei at $\sqrt{s_{NN}}= $200 GeV at RHIC. 
Cooper-Fry formulation is used to calculate the $p_T$ spectra 
for different hadrons considering the same initial conditions 
as for thermal photons. We assume that the system reaches a 
state of maximum entropy at the point of thermalization at a 
time $\tau_0$ and then follows an isentropic expansion. 
The initial entropy density $s_0$ is obtained by combining `hard' 
and `soft' contributions which are the contributions from 
binary collisions ($n_b$)  wounded nucleons ($n_w$) respectively, 
and it follows the relation~\cite{uk},
\begin{eqnarray}
s( \tau_0, \ x, \ y, \ b )= \kappa \ [ \ \alpha \ n_w(x, \ y, \ b) \ + 
\ ( \ 1 \ - \ \alpha \ ) \ n_b(x,\ y, \ b) \ ].
\label{s0}
\end{eqnarray} 
$\kappa$ and $\alpha$ (=0.75) are the constants in Eq.~\ref{s0} 
and Glauber model formulation is used to calculate the values of 
$n_b$ and $n_w$ at different points in the transverse plane for 
a particular impact parameter b. The value of $\kappa$ is obtained 
considering the initial entropy density  at x=y=b=0 is about 117 
fm$^{-3}$ when $\tau_0$ is 0.6 fm/$c$.~\cite{ksh}. We use boost 
invariant and impact parameter dependent azimuthally anisotropic 
hydrodynamics to model our system, where the  plasma experiences a 
first order phase transition at a temperature of about 164 MeV and 
the freeze-out energy density is considered at 0.075 GeV/fm$^3$. 
Thermal photon emission from the quark matter and the hadronic matter 
are obtained by integrating the rates of emission over the space time 
volume $ \rm{d^4x \ ( = \ dx \ dy  \ \tau d\tau \ d\eta )}$. Photons 
from quark matter are calculated considering a complete leading 
order photon production rate from Arnold {\it et al.}~\cite{AMY} 
and  we use the latest results by Turbide {\it et al.}~\cite{TRG} 
for thermal radiation from the hot hadronic gas. The elliptic flow 
parameter $v_2$, which directly reflects the rescattering among the 
particles produced in the collisions, is obtained using the relation,
\begin{equation}
\frac{dN(b)}{d^2p_T \, dy}=\frac{dN(b)}{2 \pi p_T \, dp_T \, dy}
\left[ \ 1 \ + \ 2 v_2(p_T,b) \ \cos(2 \phi) \ +...\right ].
\end{equation}
\subsection{Hadron spectra and $v_2$ at different $\tau_0$}
The $p_T$ spectra and $v_2(p_T)$ of several hadrons (nearly 
complete list available in the particle data book) are calculated 
at different $\tau_0$ (ranging from 0.2 to 1.0 fm/$c$, in steps 
of 0.2 fm/$c$ ) by keeping the total entropy of the system fixed.
Particle spectra and elliptic flow results for $\rho$ mesons at 
different $\tau_0$ for semi-central collision of Au nuclei are 
shown in Fig.~\ref{rho}.
\begin{figure}[t]
\begin{minipage}{15pc}
\includegraphics[width=13pc,angle=-90]{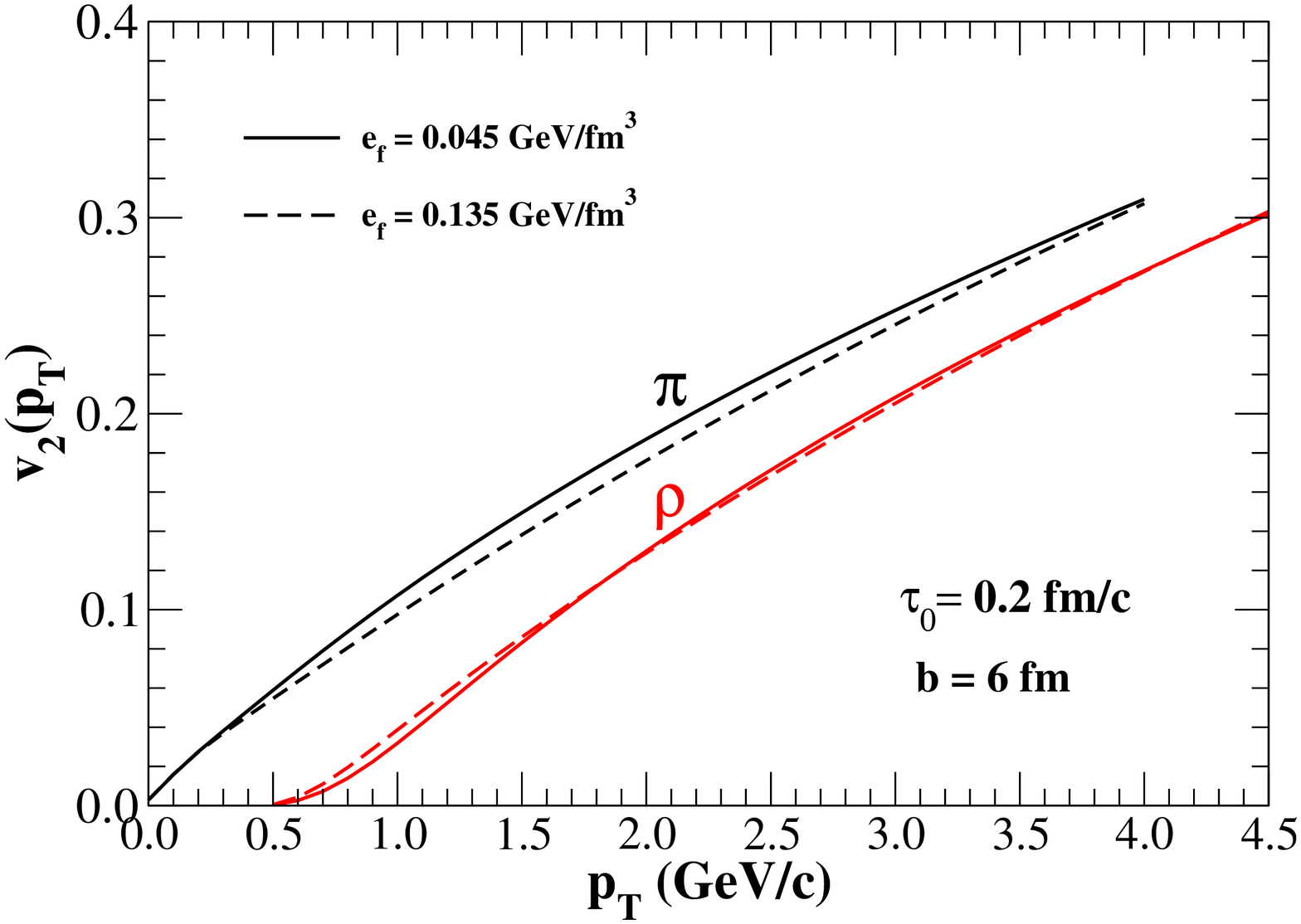}
\end{minipage}\hspace{1pc}%
\begin{minipage}{15pc}
\includegraphics[width=13pc,angle=-90]{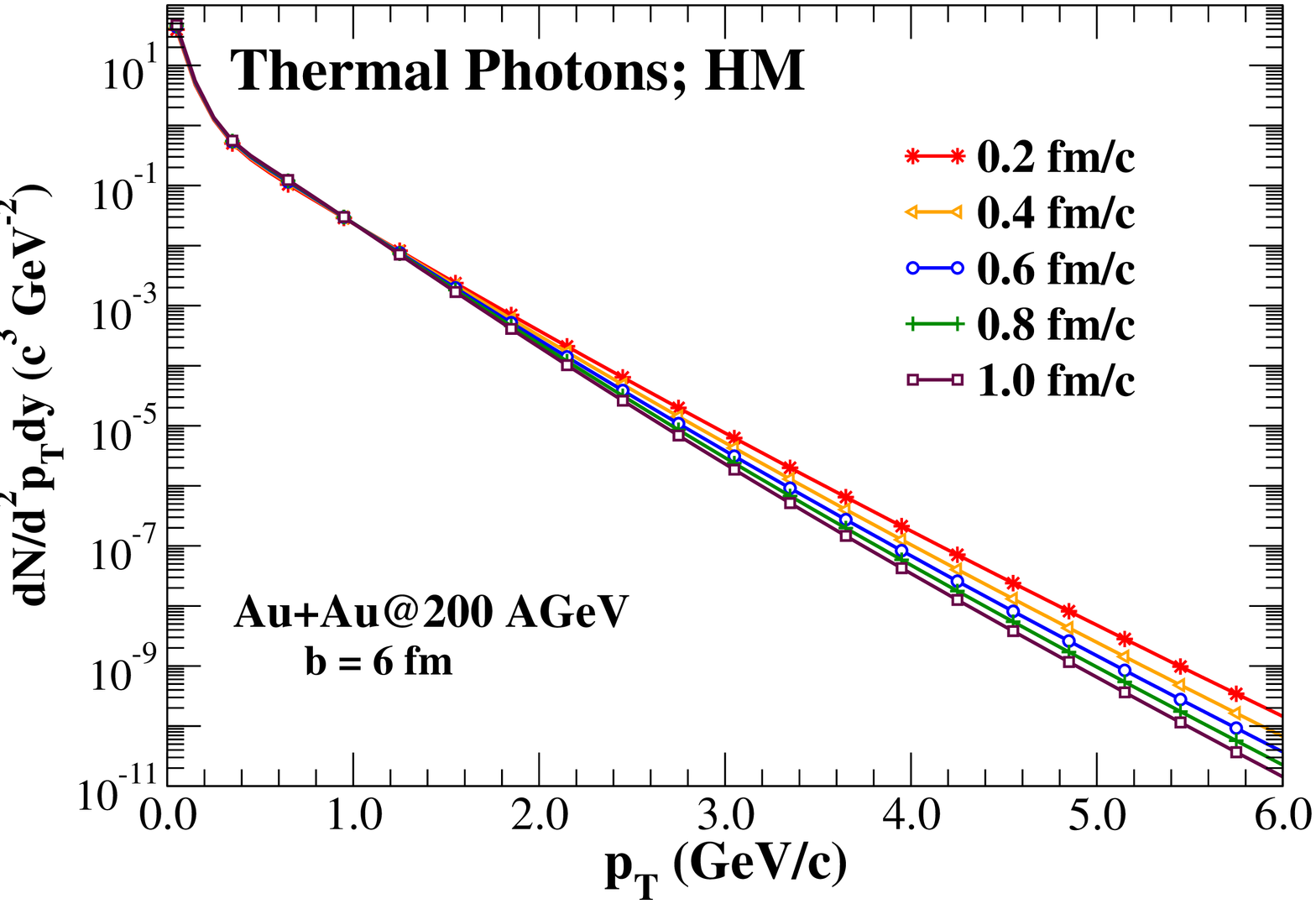}
\end{minipage} 
\caption{\label{2} [Left panel] Elliptic flow of primary 
hadrons with changing freeze-out density. [Right panel] Thermal 
photons from hadronic phase (only) at different $\tau_0$.}
\end{figure}
The emission of hadrons is mainly governed by the conditions 
of the freeze-out hyper-surface or, in particular, the 
temperature at freeze-out ($T_f \sim 120$ MeV).  We see that 
the heavier particles respond strongly to radial flow with 
changing $\tau_0$, whereas the elliptic flow results for all 
the hadrons remain almost unaffected with changing initial 
formation time of the plasma. Early start of flow drives the 
freeze-out to happen sooner and hence the overall flow does 
not change significantly with changing $\tau_0$ for them. We
find the same insensitivity to $\tau_0$  for all the hadrons in 
the particle data book. The effect of changing freeze-out conditions 
on the hadron spectra and elliptic flow are checked and resultant 
elliptic flow for $\pi$ and $\rho$ mesons are shown in left panel 
of Fig.~\ref{2}. We find that the flow parameter for hadrons
essentially acquire its final value at some larger temperature 
and does not change significantly with changing freeze-out conditions.

\subsection{Thermal photon spectra and elliptic flow at different $\tau_0$}
We see that, the thermal photon spectra from hadronic phase at 
different $\tau_0$ are almost independent of the value of  $\tau_0$ 
for $p_T < 1.5 $ GeV/$c$. Only at very large $p_T \ ( \sim 5 \ \rm{GeV}
/{\it c})$ values, the spectra at a smaller $\tau_0$ is flatter than 
that  at a larger $\tau_0$ [see right panel of Fig.~\ref{2}]. 
\begin{figure}[t]
\begin{minipage}{15pc}
\includegraphics[width=13pc,angle=-90]{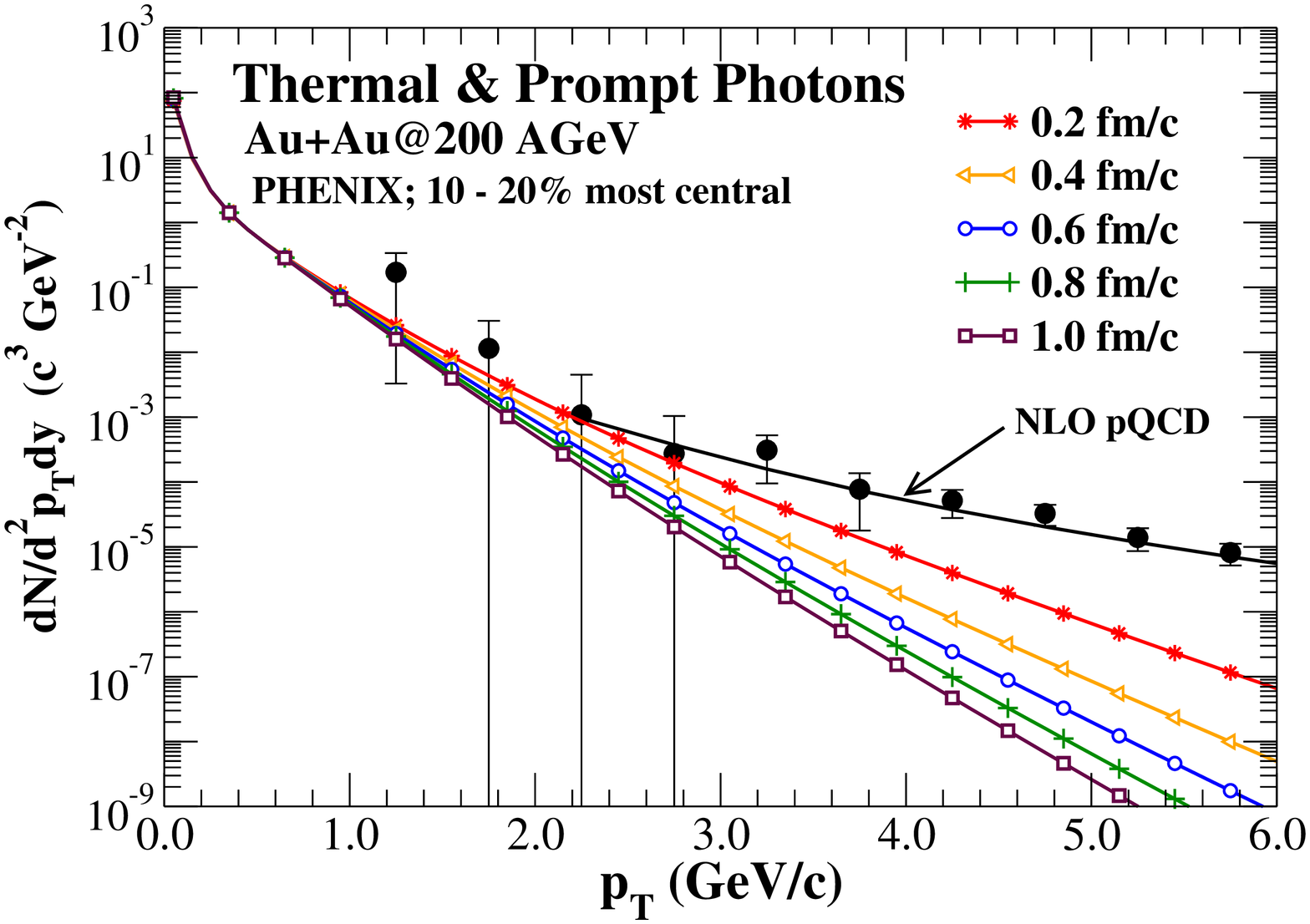}
\end{minipage}\hspace{1pc}%
\begin{minipage}{15pc}
\includegraphics[width=13pc,angle=-90]{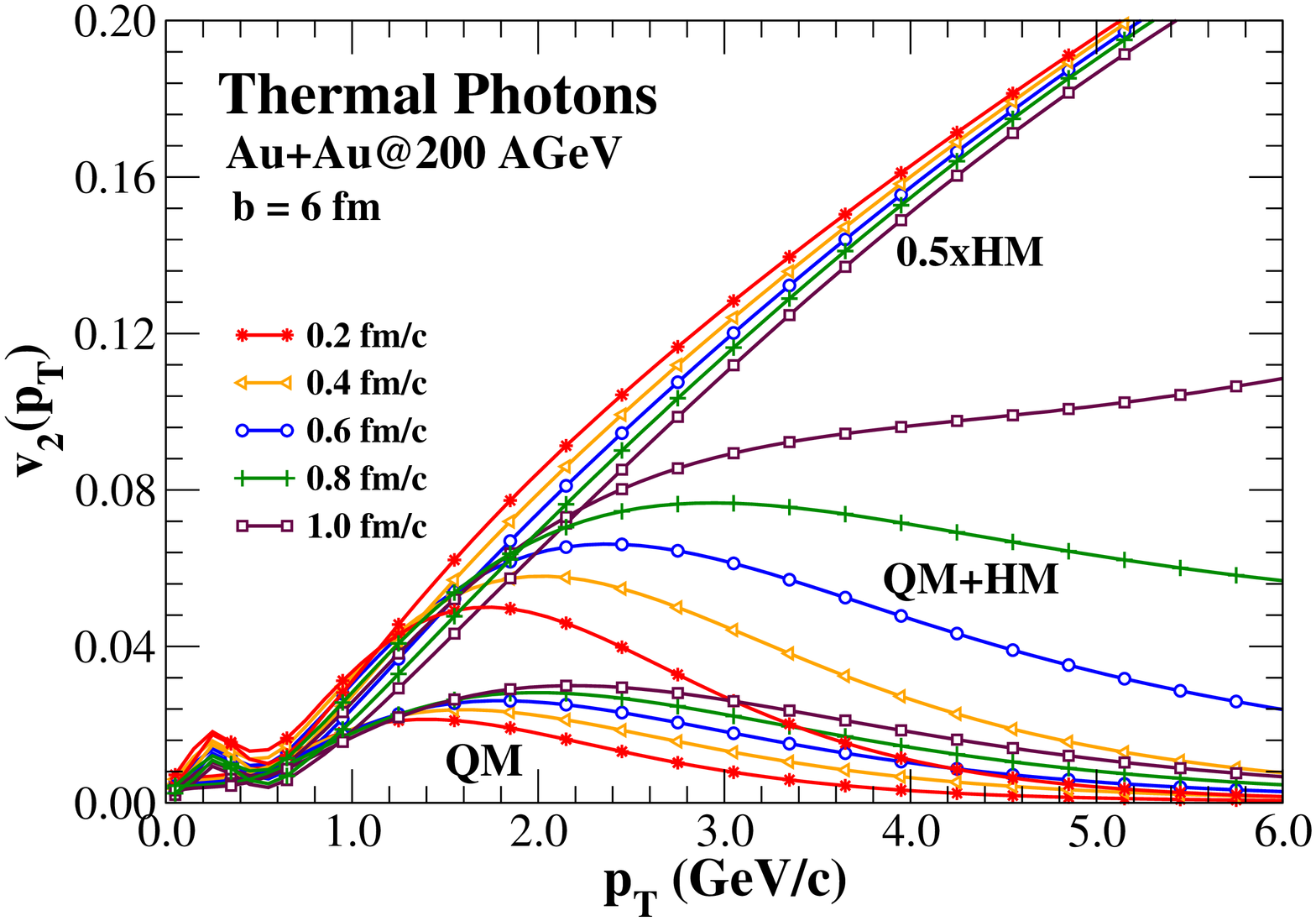}
\end{minipage} 
\caption{\label{phot} [Left panel] Thermal photon spectra at 
different formation time at RHIC. [Right panel] Differential 
elliptic flow of thermal photon for different initial time 
$\tau_0$. }
\end{figure}
We know that the total entropy of the system is related 
to the particle number density and initial parameters by the 
relation, $S(\eta) \propto \ dN/dy \propto  T_0^3 \, \tau_0$, 
where, $T_0$ is the initial temperature of the plasma at time 
$\tau_0$. With smaller values of $\tau_0$, the radiation from 
QGP phase increases significantly as the initial temperature 
of the system increases. As a result, the photon spectra at 
$\tau_0 = 0.2$ fm/$c$ is much flatter than the same at $\tau_0 
= 1.0$ fm/$c$ and the two results differ by a few order of 
magnitudes at the intermediate and high $p_T$ range. Our results 
from hydrodynamics along with prompt photon results using NLO 
pQCD~\cite{nlo} and experimental data from PHENIX~\cite{phenix} 
are shown in left panel of Fig.~\ref{phot}. Although the thermal 
photon yield from quark matter increases with smaller $\tau_0$, 
the development of elliptic flow is not substantial at very early 
times and hence the $v_2(p_T)$ from quark matter decreases 
with smaller $\tau_0$. As the $v_2$ from hadronic matter is not 
affected significantly with changing $\tau_0$, the sum $v_2$ 
decreases with $\tau_0$ beyond a  $p_T$ value of about 1.5 GeV/$c$ 
as shown in right panel of Fig.~\ref{phot}. 

In conclusion, we show that the  thermal photon $v_2$ is quite 
sensitive to the formation time of QGP and its value can be 
estimated precisely with the help of experimental determination 
of the flow parameter $v_2$, whereas the $v_2(p_T)$ for hadrons 
depends only marginally on the value of $\tau_0$.

\section*{Acknowledgments} RC would like to thank the QM09 organizing 
committee for allowing her to present the talk via telephone. Special
thanks  to David Silvermyr and Vince Cianciolo for making the 
arrangement possible.

\end{document}